\documentclass[allclo]{FBSart}
\usepackage{amsfonts}
\usepackage{amssymb}

\title{Quark-model hadron structure}
\author{A. Valcarce\instnr{1}, J. Vijande\instnr{2}, 
P. Gonz\'alez\instnr{2}, H. Garcilazo\instnr{3}}
\instlist{Departamento de F\'\i sica Fundamental,
Universidad de Salamanca, Spain
\and Departamento de F\' \i sica Te\'orica e IFIC,
Universidad de Valencia - CSIC, Spain
\and Escuela Superior de F\'\i sica y Matem\'aticas,
Instituto Polit\'ecnico Nacional, Mexico}

\runningauthor{A.\,Valcarce}
\runningtitle{Quark-model hadron structure}
\begin{document}

\maketitle
\begin{abstract}
We review some selected recent results on hadron spectroscopy
and related theoretical studies based on constituent quark models.
\end{abstract}

\section{Introduction}
Hadron spectroscopy has undergone a great renaissance in recent 
years~\cite{Ros07}. The new findings include:
low-lying excitations of $D$ and $B$ mesons, 
long-awaited missing states and new states near 
4 GeV/c$^2$ in the charmonium spectrum, 
charmed and bottom baryons, and
evidence for doubly charmed baryons.
The light hadron
sector remains also restless reporting new scalar mesons or
showing a deep theoretical interest in the high energy 
part of both the meson and baryon spectra. 

The hadron spectra should be described in terms of QCD. 
Tested to very high accuracy in the perturbative regime,
its low energy sector ({\it strong QCD}) comprising hadron 
physics, remains challenging because neither lattice nor 
perturbative methods are accurate. 
As in most other areas of physics, the keys to a qualitative
understanding of strong QCD are to identify the appropriate
degrees of freedom and the effective forces between them~\cite{Isg00}.
All roads lead to valence constituent quarks as the 
appropriate degrees of freedom~\cite{Clo07}. The effective forces,
summarizing the basic properties of QCD, must at least 
contain: a confining mechanism, a spin-spin force and a 
long-range term. This framework is what we know as the
constituent quark model. Although quark models differ in their details, 
the qualitative aspects of their spectra are determined by features 
that they share in common. These common ingredients can be used to project
expectations for new sectors~\cite{Jaf07}. 

The limitations of the quark model are as obvious as its successes. 
Nevertheless almost all hadrons can be classified as relatively simple
configurations of a few confined quarks. Nowadays, we
have the tools to deepen our understanding of strong QCD.
On one side we have at our disposal powerful numerical techniques 
imported from few-body physics: Faddeev calculations in momentum 
space~\cite{Val05}, hyperspherical harmonic expansions~\cite{Vij07}
and stochastic variational methods~\cite{Kar98}. On the other we
have an increasing number of experimental data.
In this work we review some
recent results and selected theoretical analysis on heavy
baryons, heavy mesons, and light baryons.
\begin{table}[t]
\beforetab
\begin{tabular}{ccccccccc} 
\firsthline
State & $J^P$ & $Q=s$ & $Q=c$ & $Q=b$ & $J^P$ & $Q=s$ & $Q=c$ & $Q=b$ \\ \midhline
$\Lambda (udQ)$ & $1/2^+$ & 1116,1600 & 2286,2765$^*$ & 5625 
                & $1/2^-$ & 1405,1670 & 2595          &     \\ 
                & $3/2^+$ & 1890      & 2940$^*$      &     
                & $3/2^-$ & 1520,1690 & 2628,2880$^*$ &     \\ \midhline
$\Sigma  (uuQ)$ & $1/2^+$ & 1193,1660 & 2454,2980$^*$ & 5811$^*$ 
                & $1/2^-$ & 1480,1620 & 2800$^*$      &       \\ 
                & $3/2^+$ & 1385,1840 & 2518,3077$^*$ & 5833$^*$ 
                & $3/2^-$ & 1560,1670 &               &        \\ \midhline
$\Xi     (usQ)$ & $1/2^+$ & 1318      & 2469,2577     & 5792$^*$  
                & $1/2^-$ &           & 2790          &        \\ 
                & $3/2^+$ & 1530      & 2645          &        
                & $3/2^-$ & 1820      & 2815          &        \\ \midhline
$\Omega  (ssQ)$ & $1/2^+$ &           & 2698          &        
                & $1/2^-$ &           &               &        \\ 
                & $3/2^+$ & 1672      & 2770$^*$      &        
                & $3/2^-$ &           &               &        \\ 
\lasthline
\end{tabular}
\aftertab
\captionaftertab[]{Experimentally known heavy baryons.}
\end{table}
\section{Heavy baryons}
It was already in 1985 when Bjorken 
wrote~\cite{Bjo85}: "We should strive to study triply 
charmed baryons because their excitation spectrum should be close 
to the perturbative QCD regime". The larger the number of
heavy quarks the simpler the system. In particular, doubly and triply heavy
baryons are driven only by a perturbative one-gluon exchange (OGE), while
single heavy baryons include the dynamics of light and
heavy-light quark pairs. Baryons with one, two or
three heavy quarks are the ideal laboratory to test the assumed
flavor independence of confinement~\cite{Gar07}. Table 1 resumes
the experimental situation of strange, charmed and bottom baryons.
The number of experimental data in the charm and bottom
sectors is increasing rapidly, states denoted by a star have been reported
within the last two years with quantum numbers still not determined. 
\begin{table}[b]
\beforetab
\begin{tabular}{cccccccc} 
\firsthline
M(MeV) & Full    &  No OPE   &  $\Delta E$ & 
M(MeV) & Full    &  No OPE   &  $\Delta E$ \\ \midhline
$\Sigma_b(1/2^+)$ & 5807  & 5820 & $-$13  &
$\Sigma(1/2^+)$ & 1408  & 1417 & $-$9 \\ 
$\Sigma_b(3/2^+)$ & 5829  & 5839 & $-$10 & 
$\Sigma(3/2^+)$ & 1454  & 1462 & $-$8 \\ 
$\Lambda_b(1/2^+)$ & 5624  & 5804 & $-$180  & 
$\Lambda(1/2^+)$ & 1225  & 1405 & $-$180 \\ 
$\Lambda_b(3/2^+)$ & 6388  & 6388 & $<$1  & & & &\\ 
\lasthline
\end{tabular}
\aftertab
\captionaftertab[]{Contribution of the one-pion exchange (OPE) 
to the baryon mass~\protect\cite{Gar07}.}
\end{table}

The dynamics of the light-quark pair plays a relevant role in the
strange sector, being mainly responsible for the spin splitting. 
Thus, it is expected a similar contribution for charmed and bottom baryons. 
The experimental spin splitting has been measured in the charm and bottom
sectors obtaining:
$M[\Sigma_c(3/2^+)]-M[\Lambda_c(1/2^+)]=$ 232 MeV,
$M[\Sigma_c(3/2^+)]-M[\Sigma_c(1/2^+)]=$ 64 MeV, 
$M[\Sigma_b(3/2^+)]-M[\Lambda_b(1/2^+)]=$ 209 MeV, and
$M[\Sigma_b(3/2^+)]-M[\Sigma_b(1/2^+)]=$ 22 MeV. 
These results are rather well reproduced with quark models containing
only gluons or gluons and pions in the light quark dynamics.
In Table 2 we show the contribution of pions separately,
noting how pions give the same contribution for strange and bottom 
baryons. Therefore, the contribution of gluons is diminished for models
considering pions. If we now go
to doubly charmed baryons, where pions do not contribute, the predictions
are parameter free and experiment will confirm or defeat these results 
giving hints on the underlying dynamics of the system: 
\begin{eqnarray}
M[\Xi_{cc}(3/2^+)]-M[\Xi_{cc}(1/2^+)] & =&  66 \,\, {\rm MeV}  \nonumber \\
M[\Omega_{cc}(3/2^+)]-M[\Omega_{cc}(1/2^+)]&  =&  54 \,\, {\rm MeV} \, .
\end{eqnarray}

\section{Heavy mesons}
More than thirty years after the so-called November revolution~\cite{Bjo85}, 
heavy meson spectroscopy is being severely tested by new 
experiments~\cite{Ros07}. This challenging situation arose in 
the open-charm sector
with the discovery of the $D_{sJ}^*(2317)$, the $D_{sJ}(2460)$ and the
$D_0^*(2308)$ mesons, positive parity states with masses smaller
than expectations from quark potential models.
One could say in general that the area phenomenologically 
understood in the open-charm meson spectrum extends to
states where the $q\bar q$ pair is in relative $S-$wave. In the positive
parity sector, $P-$wave states, is where the problems arise. This has been said
as an example where naive quark models are probably too naive~\cite{Clo07}. 
Out of the many explanations suggested for these states, 
the unquenching of the naive quark model has been successful~\cite{Vij05}. 
When a $q\bar q$ pair occurs in a $P-$wave but can couple to hadron 
pairs in $S-$wave the latter will distort the $q\bar q$ picture. 
In the examples mentioned above, the $0^+$ and $1^+$ $c\bar s$ states
predicted above the $DK(D^*K)$ thresholds couple to the continuum. This 
mixes $DK(D^*K)$ components in the wave function.
This idea can be easily formulated in terms of a meson wave-function
described by 
\begin{equation}
\label{mes-w}
\left|\psi\right>=\sum_{i} \alpha_i \left|q
\bar q\right>_i + \sum_{j} \beta_j \left|qq\bar q \bar q\right>_j
\label{eq-w}
\end{equation}
where $q$ stands for quark degrees of
freedom and the coefficients $\alpha_i$ and $\beta_j$ take into
account the admixture of four-quark components in the
$q \bar q$ picture.

Results for the open-charm mesons~\cite{Vij05} show that
they are easily identified with
standard $c \overline{q}$ states except for
the $D_{sJ}^*(2317)$, the $D_{sJ}(2460)$, and the $D^*_0(2308)$.
Thus, one could be tempted to interpret them as four-quark resonances
within the quark model. Other results obtained with the 
same interacting potential~\cite{Vij05} are: for 
$cn\bar s\bar n$, $(J^P,I)=(0^+,0)$ 2731 MeV,
$(0^+,1)$ 2699 MeV,
$(1^+,0)$ 2841 MeV,
$(1^+,1)$ 2793 MeV, and for
$cn\bar n\bar n$,
$(0^+,1/2)$ 2505 MeV. All of them are far above the corresponding 
strong decay threshold and therefore broad in contrast to
experiment, what rules
out a pure four-quark interpretation.

Thus, physical states may correspond to a mixing of 
two- and four-body configurations, Eq. (\ref{eq-w}). 
The results obtained are shown in Table 3
In the nonstrange sector once the mixing is considered
one obtains a state at 2241 MeV with 46\% of four-quark component 
and 53\% of $c\bar n$ pair. The lowest state, representing
the $D^*_0(2308)$, is above the isospin preserving threshold $D\pi$,
being broad as observed experimentally. 
The mixed configuration compares much better with 
the experimental data than the pure $c\bar n$ state. 
The orthogonal state appears at 2713 MeV, with
and important four-quark component. 
In the strange sector, the $D_{sJ}^*(2317)$ and the $D_{sJ}(2460)$ 
are dominantly $c\bar s$ $J=0^+$ and $J=1^+$ states, respectively,
with almost  30\% of four-quark component. Without being dominant,
this percentage is fundamental to shift the mass of the unmixed states to 
the experimental values below the $DK$ and $D^*K$ thresholds
and, therefore, they are expected to have small widths. 
\begin{table}[t]
\beforetab
\begin{tabular}{ccccccccc} 
\firsthline
\multicolumn{3}{c}{$J^P=0^+(I=0)$}    & \multicolumn{3}{c}{$J^P=1^+(I=0)$} & \multicolumn{3}{c}{$J^P=0^+(I=1/2)$} \\
\midhline
QM                  &2339   &2847  &QM					&2421  &2555  	&QM                   &2241 &2713    \\
Exp.                &2317.4&$-$  &Exp.		&2459.3&2535.3&Exp.   &2308$\pm$36&$-$\\
\midhline
P($cn\bar s\bar n$) &28   &55  &P($cn\bar s\bar n$)	&25  &$\sim 1$ 	&P($cn\bar n\bar n$)  &46        &49  \\
P($c\bar s_{1^3P}$) &71   &25  &P($c\bar s_{1^1P}$)	&74  &$\sim 1$ 	&P($c\bar n_{1P}$)    &53        &46 \\
P($c\bar s_{2^3P}$) &$\sim 1$  &20  &P($c\bar s_{1^3P}$)&$\sim 1$ &98	&P($c\bar n_{2P}$)    &$\sim 1$  &5 \\
\lasthline
\end{tabular}
\aftertab
\captionaftertab[]{Probabilities, in \%, of the wave function components 
and masses (QM), in MeV, of the open-charm mesons 
once the mixing is considered~\protect\cite{Vij05}.}
\end{table}

The above arguments have also open the discussion about the presence of compact 
four-quark states in the charmonium spectrum, 
with special emphasis on the nature of the $X(3872)$.
It is a member of an heterogeneous group, including 
the $Y(2460)$ and the recently reported $Z(4430)$,
whose properties make their identification as
traditional $q\bar q$ states unlikely. 
Although some caution is still required 
an isoscalar $J^{PC}=1^{++}$ state seems to
be the best candidate to describe the $X(3872)$ properties.
\begin{table}[b]
\beforetab
\begin{tabular}{cccccccccc} 
\firsthline
 &\multicolumn{2}{c}{CQC} &\multicolumn{2}{c}{BCN} &
 &\multicolumn{2}{c}{CQC} &\multicolumn{2}{c}{BCN} \\
$J^{PC}(K_{\rm max})$ & $E_{4q}$ & $\Delta_{E}$&
$E_{4q}$ & $\Delta_{E}$ & 
$J^{PC}(K_{\rm max})$ & $E_{4q}$ & $\Delta_{E}$&
$E_{4q}$ & $\Delta_{E}$ \\ 
\midhline
$0^{++}$ (24) & 3779 &  +34 &  3249 &  +75  &
$0^{--}$ (17) & 3791 & +108 &  3405 & +172  \\
$0^{+-}$ (22) & 4224 &  +64 &  3778 & +140  & 
$0^{-+}$ (17) & 3839 &  +94 &  3760 & +105  \\
$1^{++}$ (20) & 3786 &  +41 &  3808 & +153  &
$1^{--}$ (19) & 3969 &  +97 &  3732 &  +94  \\
$1^{+-}$ (22) & 3728 &  +45 &  3319 &  +86  & 
$1^{-+}$ (19) & 3829 &  +84 &  3331 & +157  \\
$2^{++}$ (26) & 3774 &  +29 &  3897 &  +23  & 
$2^{--}$ (21) & 4054 &  +52 &  4092 &  +52  \\
$2^{+-}$ (28) & 4214 &  +54 &  4328 &  +32  & 
$2^{-+}$ (21) & 3820 &  +75 &  3929 &  +55  \\
\lasthline
\end{tabular}
\aftertab
\captionaftertab[]{Mass, in MeV, of the different $J^{PC}$ $c\bar c n\bar n$ 
states, $E_{4q}$, calculated including up to $K_{\rm max}$ HH,
and difference with the lowest two-meson 
threshold, $\Delta_E$ \protect\cite{Vij07}.}
\end{table}

Charmonium four-quark states have been studied solving the four-body Schr\"odinger
equation using the hyperspherical harmonic (HH) formalism~\cite{Vij07}
with two standard quark-quark interaction models: one based only on
the one-gluon exchange (BCN), and the other 
containing also boson exchanges (CQC). 
As can be seen in Table 4
there appear no bound states for any set of quantum numbers, including
the suggested assignments of the $X(3872)$, $1^{++}$.
Independently of the quark-quark interaction and the quantum numbers 
considered, the system evolves to a well separated two-meson state. 
Thus, in any manner one can claim for the existence
of a bound state for the $c\bar c n \bar n$ system
unless additional ingredients either in the interaction
or in the wave function are considered. 

\section{Light baryons}
\begin{table}[b]
\beforetab
\begin{tabular}{ccc} 
\firsthline
$(K,L,Symmetry)$ & $S=1/2$ & $S=3/2$ \\ \midhline
$(0,0,[3])$ & $N(1/2^{+})(940)$ & $\Delta (3/2^{+})(1232)$ \\ 
$(2,2,[3])$ & $N(5/2^{+})(1680),N(3/2^{+})(1720)$ & 
$\Delta (7/2^{+})(1950)$ \\ 
$(4,4,[3])$ & $N(9/2^{+})(2220)$ & 
$\Delta (11/2^{+})(2420)$ \\ 
$(6,6,[3])$ & $N(13/2^{+})(\ast \ast )(2700)$ & 
$\Delta (15/2^{+})(\ast \ast )(2950)$ \\ \midhline
$(2,0,[21])$ & $N(1/2^{+})(\ast \ast \ast )(1710)$,
$\Delta (1/2^{+})(1750)$ & \\ 
$(2,2,[21])$  & $N(5/2^{+})(\ast \ast )(2000)$,
$\Delta (5/2^{+})(1905)$ & $N(7/2^{+})(\ast \ast )(1990) $ \\ 
$(4,4,[21])$ & $N(9/2^{+})(2220)$,
$\Delta (9/2^{+})(\ast \ast )(2300)$ &
$N(11/2^{+})(?)$ \\ 
$(6,6,[21])$ & $N(13/2^{+})(2700)$,
$\Delta (13/2^{+})(?)$ & $N(15/2^{+})(?)$ \\ \midhline
$(1,1,[21])$ & $N(3/2^{-})(1520),N(1/2^{-})(1535)$ & 
$N(5/2^{-})(1675)$ \\ 
& $\Delta (3/2^{-})(1700),\Delta (1/2^{-})(1620)$ & \\ 
$(3,3,[21])$ & $N(7/2^{-})(2190)$,
$\Delta (7/2^{-})(\ast )(2200)$ & $N(9/2^{-})(2250)$ \\ 
$(5,5,[21])$ & $N(11/2^{-})(\ast \ast \ast )(2600)$,
$\Delta (11/2^{-})(?)$ & $N(13/2^{-})(?)$ \\ 
$(3,3,[3])$ & $N(7/2^{-})(?)$ & $\Delta (9/2^{-})(\ast \ast )(2400)$ \\ 
$(5,5,[3])$ & $N(11/2^{-})(?)$ & $\Delta (13/2^{-})(\ast \ast )(2750)$
\\ 
\lasthline
\end{tabular}
\aftertab
\captionaftertab[]{
Dominant spatial-spin configurations for $N's$ and $\Delta's$.}
\end{table}

The high energy part of the baryonic spectrum has
been a subject of interest in the last decade, the aim being to get a better
understanding of the confinement mechanism and the hadronization process.
In particular the idea of a parity multiplet classification scheme at high excitation
energies as due to chiral symmetry was suggested some years ago \cite{Jid00}
and put in question later on \cite{Jaf05}. 

The use of a quark-quark screened confining potential 
supplemented by a minimal OGE (coulomb+hyperfine)
\cite{Gon06}, allows to obviate the
missing state problem. A correct prediction of the number and ordering of the
known $N$ and $\Delta $ resonances, up to 2.4 GeV mass 
is obtained. The unambiguous assignment of quantum numbers to the dominant
configuration of any $J^{P}$ ground and first non-radial states up to
$J=11/2$ translates into a well defined symmetry pattern.

In Table 5 we group
experimental resonances according to their dominant configuration.
States denoted by a question mark have not an assigned mass
in the Particle Data Review. 
To express the spatial part we use the 
quantum numbers $(K,L,Symmetry)$. 
$K$, defines the parity of the state, $P=(-)^{K}$, and its
centrifugal barrier energy. $L$
is the total orbital angular momentum. $Symmetry$ specifies the spatial
symmetry.
A look at the table makes manifest the underlying 
$SU(4)\otimes O(3)$ symmetry providing a $(20,L^{P})$ classification scheme,
the 20plet structure coming out naturally from the product of irreducible
quark representations: $4\otimes 4\otimes 4=20_{S}\oplus 20_{M}\oplus 20_{M}
\oplus \overline{4}$.
\begin{table}[t]
\beforetab
\begin{tabular}{ccccc} 
\firsthline
$J=7/2$ & $N(7/2^+)^\bullet$(2220) & $N(7/2^-)^\bullet$(2250) & & $
\Delta(7/2^-)^\bullet$(2400) \\ 
$J=9/2$ & $N(9/2^+)^\bullet$(2450) & $N(9/2^-)^\bullet$(2600) & $
\Delta(9/2^+)^\bullet$(2420) & $\Delta(9/2^-)^\bullet$(2650) \\ 
$J=11/2$ & $N(11/2^+)$(2450) & & & $\Delta(11/2^-)$(2650) \\ 
& $N(11/2^+)^\bullet$(2700) & $N(11/2^-)^\bullet$(2650) & $
\Delta(11/2^+)^\bullet$(2850) & $\Delta(11/2^-)^\bullet$(2750) \\ 
$J=13/2$ & & $N(13/2^-)$(2650) & $\Delta(13/2^+)$(2850) & \\ 
& $N(13/2^+)^\bullet$(2900) & & $\Delta(13/2^+)^\bullet$(2950) & \\ 
$J=15/2$ & $N(15/2^+)$(2900) & & & \\
\lasthline
\end{tabular}
\aftertab
\captionaftertab[]{Predicted $N's$ and $\Delta's $~\protect\cite{Gon06}.}
\end{table}

From the spectral pattern represented by Table 5 experimental
regularities and degeneracies for $J\geq 5/2$ ground states come out:
1.- $E_{N,\Delta }(J+2)-E_{N,\Delta }(J)\approx 400-500$ MeV;
2.- $N(J^{\pm})\approx \Delta (J^{\pm})$ for $J=\frac{4n+3}{2}$, $n=1,2...$;
3.- $N(J^{+})\approx N(J^{-})$ for $J=\frac{4n+1}{2}$, $n=1,2.$; and
4.- $(N(J),\Delta (J))^{^{\bullet }}\approx (N(J+1),\Delta (J+1))$.
The black dot denotes the first non-radial excitation.
Taking into account these rules and the symmetry pattern one can
make predictions for, until now, unknown states from 2 to 3 GeV, Table 6. 
They may serve of some help to guide future experimental searches.

\begin{acknowledge}
This work has been partially funded by MCyT
under Contract No. FPA2007-65748 and by JCyL
under Contract No. SA016A07.
\end{acknowledge}

\end{document}